\documentclass[reprint,superscriptaddress,twocolumn,showpacs]{revtex4-1}
\pdfoutput=1

\usepackage{amssymb,amsmath,amsthm}
\usepackage{dsfont}
\usepackage{fouridx}
\usepackage{graphicx,ctable,booktabs,subfigure}
\usepackage{setspace}
\usepackage[T1]{fontenc} %%bold textsc

\newcommand{\non}{\nonumber}

\newcommand{\bra}[1]{ \langle #1 |}
\newcommand{\ket}[1]{ |#1 \rangle}
\newcommand{\proj}[1]{| #1  \rangle  \langle #1 |}

\newcommand{\vac}{\textrm{vac}}

\DeclareMathOperator{\tr}{tr}

\newcommand{\td}{\textrm{d}}
\newcommand{\measure}[2]{ {\td^{#2} #1} \ }
\newcommand{\fracmeasure}[3]{\frac{\td^{#3} #1}{#2} \ }

\newcommand{\veps}{\varepsilon}

\newcommand{\cE}{\mathcal{E}}

\newcommand{\cI}{\mathcal{I}}

\newcommand{\scnl}{{\textsc{nl}}}
\newcommand{\scl}{{\textsc{l}}}

\newcommand{\tin}{\textrm{in}}
\newcommand{\tprim}{\textrm{prim}}
\newcommand{\tcascprim}{\textrm{c-prim}}

\newcommand{\logical}{\textrm{L}}

\newcommand{\cav}{\textrm{cav}}
\newcommand{\atom}{\textrm{atom}}
\newcommand{\ff}{\textrm{field}}
\newcommand{\ffcav}{\textrm{field-cav}}
\newcommand{\cavatom}{\textrm{cav-atom}}

\newcommand{\spont}{\gamma_{\textrm{3D}}}

\newcommand{\re}[1]{\textrm{Re}#1}
\newcommand{\im}[1]{\textrm{Im}#1}

\newcommand{\wt}[1]{\widetilde{#1}}
\newcommand{\Eqref}[1]{\textrm{Eq.}\,\eqref{#1}}
\newcommand{\Equationref}[1]{\textrm{Equation}\,\eqref{#1}}

\newcommand{\Figref}[1]{\textrm{Fig.}\,\ref{#1}}
\newcommand{\Figureref}[1]{\textrm{Figure}\,\ref{#1}}

\newcommand{\Secref}[1]{\textrm{Sec.}\,\ref{#1}}

\newcommand{\Slash}[1]{#1\kern-0.45em/}

\newcommand{\pos}{0}
\newcommand{\scatteringmatrix}{\mathcal{S}}

\newcommand{\Ham}{\mathcal{H}}
\newcommand{\scatteringmatrixnl}{\fourIdx{}{}{}{  \scnl  }{\scatteringmatrix}}

\newcommand{\scatteringmatrixld}{\fourIdx{}{}{\dagger}{  \scl  }{\scatteringmatrix}}

\graphicspath{{./Figures/}}

\begin{document}

%\preprint{}

\title{Deterministic and cascadable conditional phase gate for photonic qubits}

% \email, \thanks, \homepage, \altaffiliation all apply to the current author. Explanatory text should go in the []'s, actual e-mail

% \affiliation command. The \affiliation command should follow the other information.
\author{Christopher Chudzicki}
\author{Isaac L. Chuang}
\affiliation{Department of Physics, Massachusetts Institute of Technology, Cambridge, Massachusetts 02139, USA}
\affiliation{Research Laboratory of Electronics, Massachusetts Institute of Technology, Cambridge, Massachusetts 02139, USA} 
\author{Jeffrey H. Shapiro}
\affiliation{Research Laboratory of Electronics, Massachusetts Institute of Technology, Cambridge, Massachusetts 02139, USA}
%\thanks{}


\date{\today}

\begin{abstract}
Previous analyses of conditional $\phi_{\scnl}$-phase gates for photonic qubits that treat cross-phase modulation (XPM) in a causal, multimode, quantum field setting suggest that a large ($\sim$$\pi$\,rad) nonlinear phase shift is always accompanied by fidelity-degrading noise [J. H. Shapiro, Phys. Rev. A  \textbf{73}, 062305 (2006); J. Gea-Banacloche, Phys. Rev. A \textbf{81}, 043823 (2010)]. Using an atomic $\vee$-system to model an XPM medium, we present a conditional phase gate that, for sufficiently small nonzero $\phi_{\scnl}$, has high fidelity. The gate is made cascadable by using using a special measurement, principal mode projection, to exploit the quantum Zeno effect and preclude the accumulation of fidelity-degrading departures from the principal-mode Hilbert space when both control and target photons illuminate the gate.
\end{abstract}

%\keywords{}

\maketitle

\section{Introduction}

In optical quantum logic, qubit states are usually encoded using the presence or absence of a single photon in one of the many modes of the quantum electromagnetic field. We refer to this special information-carrying mode as the  \emph{principal mode}. Logic gates can be high-fidelity only if they map input principal modes to output principal modes. Gates can be cascaded successfully if the input and output principal modes are the same. In either the dual-rail  or polarization architectures, high-fidelity, cascadable single qubit-gates can be readily implemented using linear optics (beam splitters and phase-shifters). A significant challenge to implementing optical quantum information processing is the faithful realization of a deterministic and cascadable universal two-qubit photonic logic gate.

Cross-phase modulation (XPM)---a nonlinear process in which one electric field affects the refractive index seen by another---has often been proposed \cite{GJMilburn1989,Turchette1995,Chuang1995,Lukin2000} as a nonlinear optical process that might be used to construct such a universal gate, the conditional $\pi$-phase gate. (Other, fundamentally different photonic two-qubit gates have been designed, e.g., \cite{Kimble2004,Koshino2010}, which involve only single-photon+atom interactions; such gates will not be discussed here.) While a single-mode analysis of XPM-based  gates is encouraging, in recent years multimode efforts \cite{JHS2006,JHS2007,JGB2010}  that treat photons as excitations of a quantum field with continuously many degrees of freedom have been somewhat more foreboding. 

In \cite{JHS2006,JHS2007} the problem was studied using a quantized version of the solution to the classical coupled mode equations for XPM. It was shown that, within this model of quantum XPM, noise terms necessary to preserve commutation relations prevent the high-fidelity operation of a conditional $\phi_{\scnl}$-phase gate when $\phi_{\scnl} \sim \pi$. Similar fidelity-degrading noise arose in the work of Gea-Banacloche \cite{JGB2010}, whose treatment of XPM was based on a Hamiltonian describing an effective field-field interaction appropriate to a medium exhibiting electromagnetically induced transparency (EIT). In this analysis the difficulties were attributed, at least partially, to spontaneous emission.

It was recently demonstrated \cite{Koshino2009} that these problems can be circumvented by encoding qubit states in resonant, temporally-entangled (highly-bunched) biphoton pairs and using an atomic $\vee$-system to realize a Kerr medium. However, this approach is not scalable: while it may be a reasonable way to implement a conditional $\pi$-phase gate on exactly one pair of qubits, implementing this gate on any pair of $n$ qubits would require encoding $n$-qubit states in $n$-photon packets every pair of which is temporally entangled. 

The prospects for achieving a high-fidelity conditional $\phi_{\scnl}$-phase gate for photonic qubits with $\phi_{\scnl} \sim \pi$ thus seem rather dim. Still, semiclassical analyses have shown that several media, such as those supporting the EIT-based giant Kerr effect \cite{Schmidt1996,Fleisch2005}, possess $\chi^{(3)}$ nonlinearities whose real part (responsible for the XPM phase shift), though small, is large in comparison to the rate at which various fidelity-degrading absorption processes occur. With this in mind, we address in the present work the following question. Can a high-fidelity conditional $\phi_{\scnl}$-phase gate be constructed for \emph{small} $\phi_{\scnl}$, and could these gates be cascaded to yield a significant nonlinear phase shift with high-fidelity?

We show that, indeed, a conditional phase gate can be constructed with small nonlinear phase shift such that the error probability (infidelity) $|\veps|^2$ is even smaller, $|\veps|^2 \ll \phi_{\scnl} \ll 1$. Cascading these gates, however, is nontrivial.  The error, which results from a slight deformation of the principal modes, can be coherently amplified as the gate is cascaded, preventing the straightforward construction of a conditional $\pi$-phase gate. This difficulty can be avoided by performing a measurement after each primitive conditional $\phi_{\scnl}$-phase gate that projects onto the principal mode subspace, exploiting  the quantum Zeno effect as an error-preventing mechanism \cite{Vaidman1996,Erez2004,Paz2011}. For a particular choice of principal modes, we suggest one way that such a measurement could be realized.

In deriving these results, we start from a Hamiltonian describing the interaction of two quantum optical fields with a three-level $\vee$-atom. While the nonlinearities present in a $\vee$-atom are not as strong as those in, for example, the giant Kerr effect \cite{Schmidt1996}, the $\vee$-system is simple enough that it yields readily to an analysis in terms of quantum fields. After solving for the evolution of our system in the one- and two-photon subspace, we investigate fidelity and cascadability.

\section{The Fields and Their Interaction}
In this section we describe our encoding of qubit states in one-dimensional quantum fields, then consider how these fields evolve when interacting with an optical cavity containing an atomic $\vee$-system. Following the approach used in \cite{Koshino2009,Law2010,Koshino2010}, this is described by a Hamiltonian $\Ham_{\scnl}$ for the fields+cavity+atom system. The Hamiltonian-based approach we use is essentially the basis for an alternative description in terms of the input-output formalism \cite{Gardiner1985}.

The atomic system mediates an XPM-like interaction that is a central component in the conditional phase gates discussed later. Determining the nonlinear phase shift and error induced by the atomic interaction will be of the utmost importance in evaluating these gates. To this end,  one- and two-photon propagators for this system \cite{Koshino2009} are introduced.

\subsection{Qubit Encoding}

In our gate, qubit states are encoded  using two quasi-monochromatic, positive-frequency, photon-units, optical fields $h_z(\tau)$ and $v_z(\tau)$ \cite{JHS2009} (for convenience,   $\tau \equiv ct$ is used to measure time). We take $+z$ as the propagation direction, and ignore the transverse character of these fields throughout. The horizontally polarized field $h_z(\tau)$ and the vertically polarized field $v_z(\tau)$ are independent, and have nontrivial commutator $[h_z(\tau),h_{z'}(\tau)^{\dagger}]=[v_z(\tau),v_{z'}(\tau)^{\dagger}]=\delta(z-z')$.  

Logical qubit states are encoded as excitations of two $\emph{principal modes}$ $h$ and $v$, defined by 
\begin{align}
h^{\dagger} \equiv \int \measure{z}{} \psi(z) h_z^{\dagger}, && v^{\dagger} \equiv \int \measure{z}{} \psi(z) v_z^{\dagger},
\end{align}
where operators without explicit time dependence are in the Schr\"{o}dinger picture. With the normalization $\int \measure{z}{} |\psi(z)|^2 = 1$, $h^{\dagger}$ and $v^{\dagger}$ are interpreted, respectively as creating horizontally and vertically polarized photons  with wavefunction $\psi(z)$. We refer to all modes orthogonal to $h$ and $v$ as \emph{auxiliary}, or \emph{bath}, modes, and assume that the auxiliary modes are initially unexcited. In this case, the correspondence between logical qubit states and field states reads
\begin{subequations}
\label{eq:encoding}
\begin{align}
\ket{00}_{\logical} \quad \leftrightarrow \quad & \ket{\vac} \\
\ket{01}_{\logical} \quad \leftrightarrow \quad & \ket{H} \equiv  h^{\dagger} \ket{\vac} \\
\ket{10}_{\logical} \quad \leftrightarrow \quad & \ket{V} \equiv  v^{\dagger}\ket{\vac} \\
\ket{11}_{\logical} \quad \leftrightarrow \quad & \ket{HV} \equiv  v^{\dagger} h^{\dagger} \ket{\vac},
\end{align}
\end{subequations}
%The above could be written as one equation.
where $\ket{\vac}$ is the multimode vacuum. \Equationref{eq:encoding} could describe either a dual-rail or polarization encoding, where fields not participating in our gate have been dropped for convenience.

\subsection{Qubit Evolution and Interaction Hamiltonian}

\begin{figure}[tb]
\begin{center}
\subfigure[]{\label{fig:onesidedcavity}
\includegraphics[scale=0.25]{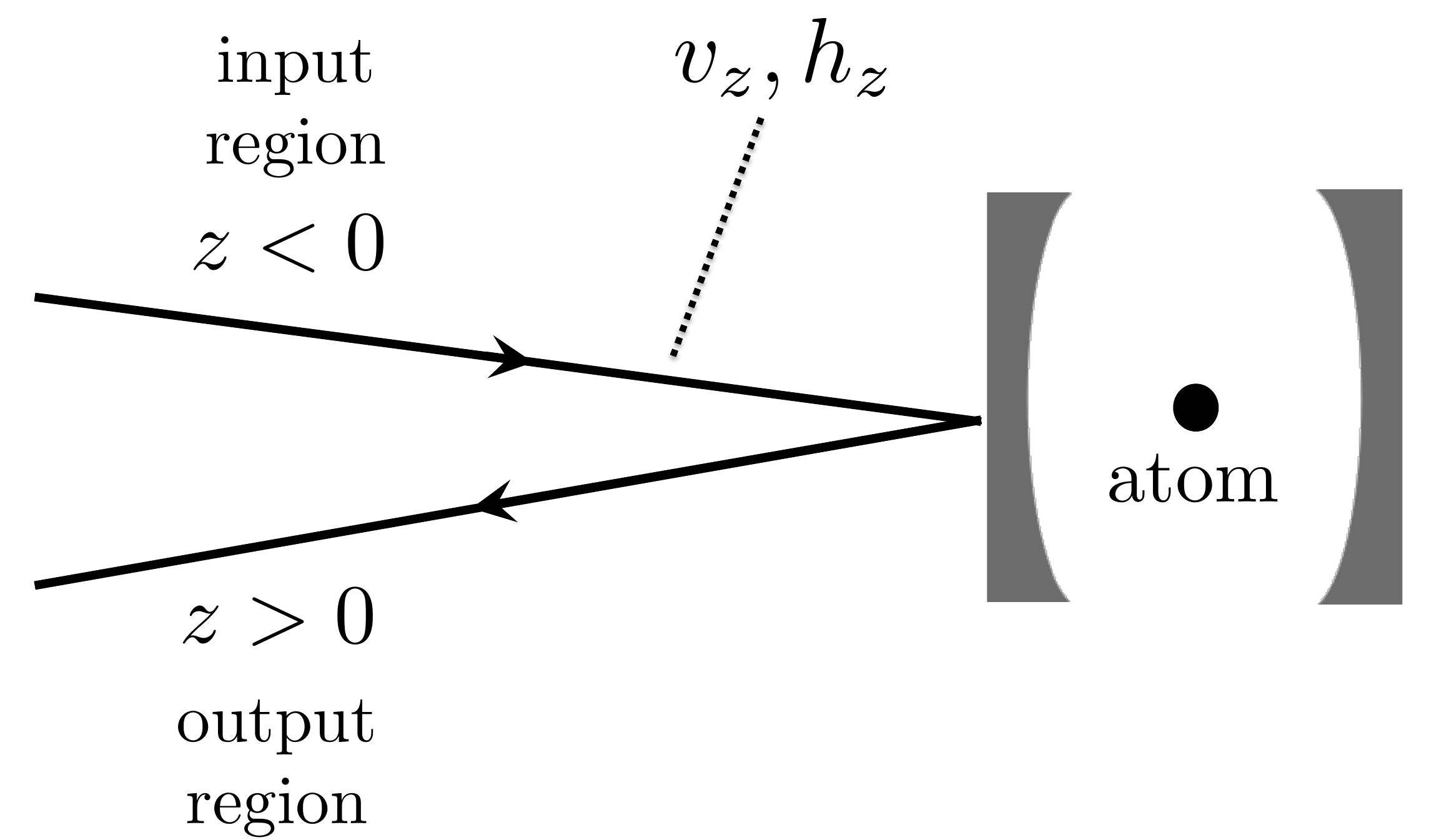} }
\subfigure[]{\label{fig:Vatom}
\includegraphics[scale=0.25]{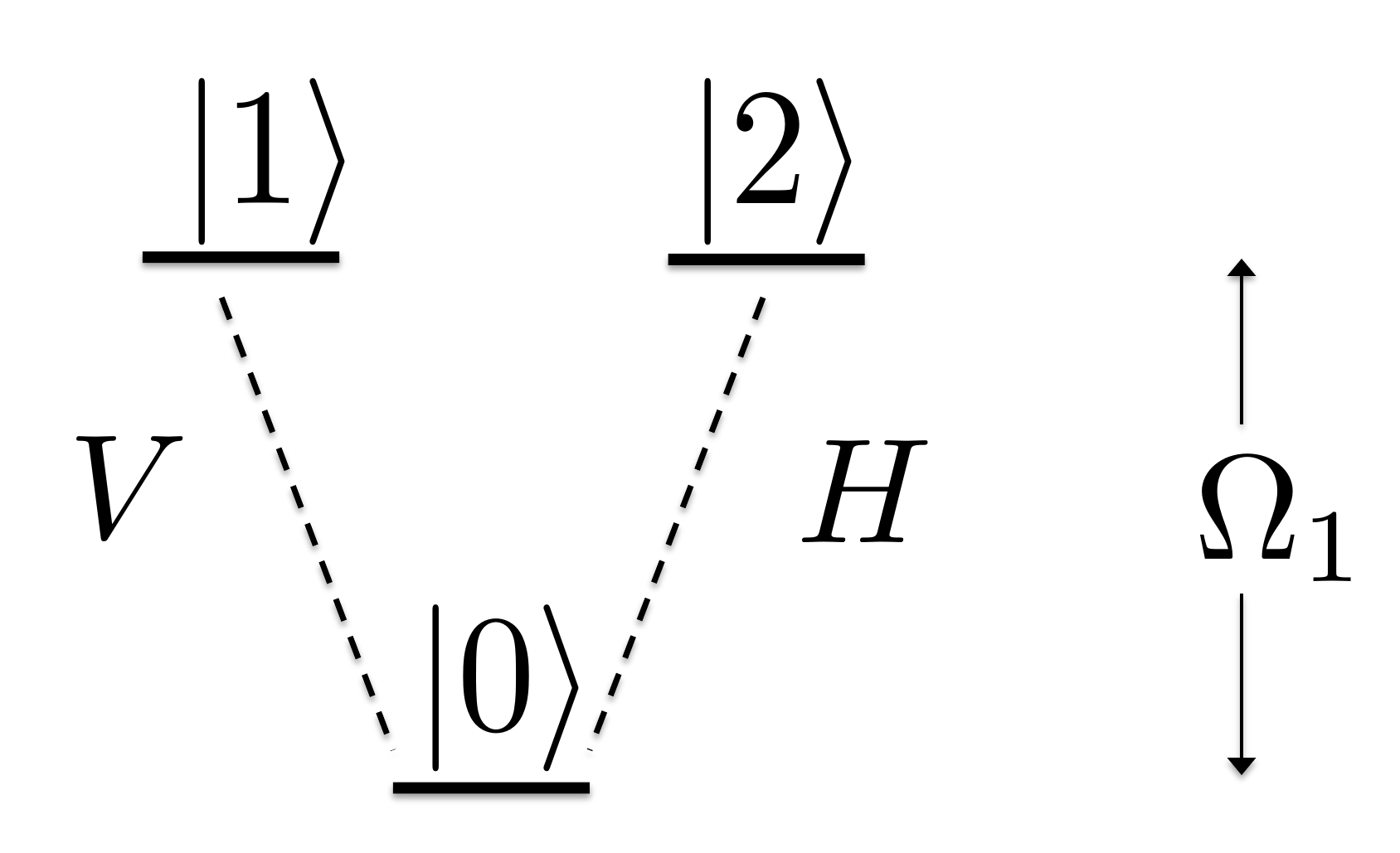} } 
\caption{(a): The external fields $v_z(\tau)$ and $h_z(\tau)$ interact with an atom placed within a one-sided cavity at position $z=0$. (b) Three-level atom used as an XPM medium. Vertical light ($V$) drives the $0 \leftrightarrow 1$ transition, while horizontal light ($H$) drives the $0 \leftrightarrow 2$ transition. In the lossy-cavity regime, $\spont \ll g \ll \kappa$, the cavity fields can be adiabatically eliminated, yielding an effective coupling directly between the external fields and the $\vee$-system at strengths $\Gamma_{H(V)} = 4 g_{H(V)}^2/\kappa_{H(V)}$. }
\label{fig:field+atom}
\end{center}
\end{figure}

At the input to our gate, the fields are prepared in some  superposition $\ket{ \psi_{\tin} }$  of the basis states in \Eqref{eq:encoding}. This state is localized in a noninteracting input region (\Figref{fig:onesidedcavity}). It then propagates in the $+z$ direction toward  a region where both fields interact, evolving under a nonlinear total Hamiltonian $\Ham_{\scnl}$ which couples these fields to a three-level atomic $\vee$-system. (Here ``nonlinear'' means that the total Hamiltonian $\Ham_{\scnl}$ generates nonlinear Heisenberg equations of motion, a necessary condition for $\Ham_{\scnl}$ to effect a two-qubit gate that does not factorize into a product of one-qubit gates. )  A long time later, the atom has returned to its ground state and the photonic qubits are in a state $\ket{\psi_1}$ which is localized in a noninteracting output region.  Working in an interaction picture with respect to the free-field Hamiltonian $\Ham_{\ff}$, the scattering matrix connects the states $\ket{\psi_{\tin}}$ and $\ket{\psi_{1}}$:
\begin{align}
\label{eq:psi1}
\ket{\psi_1} = \scatteringmatrixnl \ket{ \psi_{\tin} }, && \scatteringmatrix_{\scnl} \equiv \lim_{\tau \rightarrow \infty} e^{i \Ham_{\ff} \tau/\hbar c} e^{-i \Ham_{\scnl} \tau / \hbar c}.
\end{align}

In interacting with the atom, a single horizontally polarized (vertically polarized) photon acquires a phase shift $\phi_H$ ($\phi_V$), and may undergo some amount of pulse deformation. When both a horizontal and a vertical photon are incident upon the atom at the same time, however, the presence of the horizontal photon frustrates the interaction of the vertical photon with the atom, and vice versa, i.e., the atom cannot absorb both photons simultaneously. As a result, the pair of photons picks up an extra phase shift $\phi_{\scnl}$. In this way, the $\vee$-system models a Kerr medium, and can be used to construct conditional phase gates.

To describe this interaction, we use the same Hamiltonian $\Ham_{\scnl}$ as in \cite{Koshino2009}: both fields $h_z$ and $v_z$ couple to a one-sided cavity containing an atomic $\vee$-system, whose level structure is shown in \Figref{fig:Vatom}. For $z<0$ these fields are interpreted as propagating toward the cavity, while for $z>0$ they are interpreted as propagating away from the cavity (\Figref{fig:onesidedcavity}).  All cavity modes are ignored, except a horizontally polarized mode $a_H$ and a vertically polarized mode $a_V$, both of which are resonant with the atomic transitions at frequency $\Omega_1$. The total Hamiltonian $\Ham_{\scnl}= \Ham_0+\Ham_{\ffcav}+\Ham_{\cavatom}$ is the sum of a noninteracting Hamiltonian $\Ham_0$ and two interaction pieces.  In terms of $k$-space field operators $\wt{v}_k \equiv  \int \measure{z}{} v_z e^{-i k z}$ and $\wt{h}_k \equiv  \int \measure{z}{} k_z e^{-i k z}$, which annihilate photons with definite frequency, the noninteracting Hamiltonian $\Ham_0$ is
\begin{subequations}
\label{eq:H0}
\begin{align}
\Ham_0 = & \Ham_{\ff} + \Ham_{\cav} + \Ham_{\atom}, \\
 \label{eq:H0ff} \Ham_{\ff} = &  \int \fracmeasure{k}{2\pi}{} \hbar \omega_k \left( \wt{h}_k^{\dagger} \wt{h}_k + \wt{v}_k^{\dagger} \wt{v}_k \right), \\
\Ham_{\cav} = \ & \hbar \Omega_1 c (a_V^{\dagger} a_V+ a_H^{\dagger} a_H  ), \\
\Ham_{\atom} = \ & \hbar \Omega_1 c ( \sigma_{11} +\sigma_{22}),
\end{align}
\end{subequations}
wherein $\sigma_{mn} \equiv \ket{m}\bra{n}$ and $\omega_k = c k$ \footnote{ Because we consider a one-sided cavity, we should really only integrate over $k>0$ in \Eqref{eq:H0ff}. By using a truly linear dispersion relation, $\omega_k = ck$,  we ensure that the unphysical, negative $k$ modes are so far-detuned from the atomic system as to be irrelevant.}. Under $\Ham_0$, the Heisenberg-picture field operators propagate towards $+\infty$, e.g.,  $e^{-i \Ham_{0} \tau/\hbar c} h_z e^{i \Ham_{0} \tau/\hbar c} = h_{z-\tau}$. 

 The interactions between the cavity, free-field, and atom are taken within the rotating wave approximation, so that the total Hamiltonian $\Ham_{\scnl}$ is:
\begin{subequations}
\label{eq:Hfull}
\begin{align}
\Ham_{\scnl}^{} = & \Ham_0 +  \Ham_{\ffcav}+\Ham_{\cavatom} \\
\Ham_{\ffcav} = & \ i \hbar c \kappa_V^{1/2} ( v_{\pos} a_V^{\dagger} - v_{\pos}^{\dagger} a_V ), \non \\
&+ i \hbar c \kappa_H^{1/2} ( h_{\pos} a_H^{\dagger} - h_{\pos}^{\dagger} a_H ) \\
\Ham_{\cavatom} = & \ i \hbar c g_V (  a_V \sigma_{10} -  a_V^{\dagger} \sigma_{01} ), \non \\
&+ i \hbar c g_H ( a_H \sigma_{20} - a_H^{\dagger} \sigma_{02} ).
\end{align}
\end{subequations}
We have taken $z=0$ as the cavity's position.

As in \cite{Koshino2009}, we consider the lossy-cavity regime in which the cavity decay rates $\kappa$, cavity-atom couplings $g$, and rate $\spont$ of spontaneous emission into free space satisfy $\kappa \gg g \gg \spont$. In this regime cavity decay dominates spontaneous emission, and cavity operators can be adiabatically eliminated in favor of the external field \cite{Koshino2009,Kojima2003}. The dynamics of the atom+external field system are then identical (up to an inconsequential phase shift resulting from reflection off the one-sided cavity's perfect mirror) to those generated by an effective Hamiltonian $\Ham_{\scnl}'$ in which the fields are directly coupled to the atom, 
\begin{align}
\label{eq:Hnl}
\Ham_{\scnl}' = \Ham_0 & +  i \hbar c  \Gamma_H^{1/2} \left( v_{\pos} \sigma_{10} - v_{\pos}^{\dagger} \sigma_{01} \right)  \non \\
& + i \hbar c \Gamma_V^{1/2} \left( h_{\pos} \sigma_{20} - h_{\pos}^{\dagger} \sigma_{02} \right),
\end{align}
where the effective coupling is $\Gamma_{H(V)} = 4 g_{H(V)}^2/\kappa_{H(V)}$. In this paper, dynamics are  derived exclusively from the effective Hamiltonian, $\Ham_{\scnl}'$. To ensure that the gate treats both qubits symmetrically, we will later set $\Gamma_H=\Gamma_V$, but temporarily retain subscripts for pedagogical clarity. 
%%%Note to self: I copied the phrase pedagogical clarity from somewhere. Is that ok?

\subsection{Evolution of One- and Two-Photon States}
To determine how the one- and two-photon states in \Eqref{eq:encoding} that encode the computational basis evolve under the scattering matrix $\scatteringmatrix_{\scnl}$, it suffices to know the one- and two-photon propagators (the vacuum state evolves trivially). Labeling states $\ket{\textrm{atom;field}}$, these are
\begin{subequations}
\begin{align}
G_{H}(x,y) \equiv & \bra{0;\vac} h_x \scatteringmatrixnl h_y^{\dagger} \ket{0;\vac}, \\
G_{V}(x,y) \equiv & \bra{0;\vac} v_x \scatteringmatrixnl v_y^{\dagger} \ket{0;\vac}, \\
G_{HV}(x_H,x_V,y_H,y_V) \equiv & \bra{0;\vac} h_{x_H} v_{x_V} \scatteringmatrixnl h_{y_H}^{\dagger} v_{y_V}^{\dagger} \ket{0;\vac}.
\end{align}
\end{subequations}
 The time-dependent propagators---matrix elements of $e^{-i \Ham_{\scnl}' \tau/\hbar c}$ instead of $\scatteringmatrixnl$---are given in \cite{Koshino2009}.

 The single photon propagator $G_{H}(x,y)$ gives the long-time, interaction-picture amplitude for a photon initially at position $y$ to propagate to position $x$. We will always assume that $y<0$ so that every photon can interact with the atom, located at the origin. In this case,
\begin{align}
\label{eq:gH}
G_{H}(x,y) = & \delta(x-y) - \Gamma_H^{1/2} R_H(y-x).
\end{align}
where
\begin{align}
\Gamma_H^{-1/2} R_H(\tau) \equiv & \theta(\tau) \bra{0;\vac} e^{-i \Ham_{\scnl}' \tau/\hbar c} \ket{1;\vac}
\non \\
= &  \theta(\tau) e^{-(i \Omega_1   + \Gamma_H/2) \tau}.
\end{align}
is the amplitude for the atom, excited by a horizontally polarized impulse at time zero, to still be excited a time $\tau$ later. Here $\theta(\tau)$ is the Heaviside step function, equal to 1 for $\tau>0$ and 0 for $\tau<0$. The Fourier-space propagator $\wt{G}_{H}(k,q) \equiv  \bra{0;\vac} \wt{h}_k \scatteringmatrix \wt{h}_q^{\dagger} \ket{0;\vac}$ is also useful. Using \Eqref{eq:gH}, it is
\begin{align}
\wt{G}_{H}(k,q) = & \int \measure{x}{} \measure{y}{} G_H(x,y) e^{i q y - i k x} \non \\
\label{eq:kspaceprop}= & 2\pi \delta(k-q) \frac{k-\Omega_1-i \Gamma_H/2}{k-\Omega_1+i \Gamma_H/2}.
\end{align}
Analogous results hold for $G_{V}(x,y)$.

If the atomic system were linear, it could absorb multiple photons before emitting any. In this case, the two-photon propagator $G_{HV}(x_H,x_V,y_H,y_V)$ would just be a product of single photon propagators. Instead, it is
\begin{align}
\label{eq:twophotonpropagator}
G_{HV}(x_H, & x_V,y_H,y_V) =   \non \\ 
 & G_{H}(x,y) G_{V}(x,y) \non \\
& - \Gamma_H^{1/2} \Gamma_V^{1/2} R_H(y_H-x_H) R_V(y_V-x_V)  \non \\
& \times \theta( \min[ y_H,y_V] - \max[x_H,x_V] ).
\end{align}
Here the second piece removes from $G_{H}(x_H,y_H) G_{V}(x_V,y_V)$  exactly those terms that correspond to two absorptions before any emissions. This causes two-photon output states to be antibunched. 

The corresponding two-photon Fourier-space propagator is
\begin{align}
\label{eq:kspace2photon}
 \wt{G}_{HV}(k_H,k_V,q_H,q_V)  = & \wt{G}_H(k_H,q_H) \wt{G}_V(k_V,q_V) \non \\
 + & i \Gamma_H \Gamma_V  (2 \pi) \delta(k_H+k_V-q_H-q_V) 
\non \\
\times & \frac{1}{\wt{\delta}_{k_H}^{(H)}}  \frac{1}{\wt{\delta}_{k_V}^{(V)}} \left(  \frac{1}{\wt{\delta}_{q_H}^{(H)} }+  \frac{1}{\wt{\delta}_{q_V}^{(V)}} \right)
\end{align}
wherein $\wt{\delta}_k^{(H/V)} \equiv k - \Omega_0 + i \Gamma_{H/V}/2$. The Fourier-space propagators $\wt{G}_H,\wt{G}_V$, and$ \wt{G}_{HV}$ enable the gate fidelity calculations reported in \Secref{sec:gatefid}

\section{A Primitive Conditional Phase Gate}
In this section we describe a conditional phase gate based on the interaction $\scatteringmatrixnl$ described above. We first discuss how the unnecessary and undesirable linear evolution can be removed. We then consider the fidelity of this primitive (non-cascaded) gate with an ideal conditional phase gate. 

\subsection{Removing Linear Evolution}
\label{sec:removinglinearevolution}

In interacting with the atomic $\vee$-system, both the one- and two-photon states that encode the computational basis (\Eqref{eq:encoding}) evolve nontrivially:
\begin{subequations}
\begin{align}
\ket{H} \xrightarrow{\scatteringmatrixnl} & (1 - |\veps_H|^2)^{1/2} e^{i \phi_H} \ket{H} + \veps_H \ket{e_H}, \\
\ket{V} \xrightarrow{\scatteringmatrixnl}  & (1 - |\veps_V|^2)^{1/2} e^{i \phi_V} \ket{V} + \veps_V \ket{e_V}, \\
\ket{HV} \xrightarrow{\scatteringmatrixnl}  & (1 - |\veps_{HV} |^2)^{1/2} e^{i (\phi_H+\phi_V + \phi_{\scnl} )} \ket{HV} \non \\
& + \veps_{HV} \ket{e_{HV} }.
\end{align}
\end{subequations}
Here all kets are normalized, $\{ \phi_H, \phi_V\}$ are the single-photon (linear) phase shifts, and the various $\veps$-terms represent errors that occur because of photons evolving out of the principal modes. 

The linear phase shifts $\{ \phi_H, \phi_V\}$ are not only irrelevant to the construction of conditional logic gates, but come also with some amount of fidelity-degrading evolution out of the principal mode subspace  In order to build high-fidelity gates, it would be useful to remove completely the linear evolution that causes these effects. Removing linear evolution is also theoretically appealing because it allows one to study the fundamental limitations of the $\vee$-system's capacity for quantum XPM.

Formally, linear evolution is removed by evolving backward in time under a linearized Hamiltonian $H_{\scl}(\Omega_1)$ in which the atomic lowering operators $\sigma_{01}$ and $\sigma_{02}$ are replaced by independent harmonic oscillator annihilation operators $b_V$ and $b_H$ (\emph{c.f.} \Eqref{eq:Hnl}):
\begin{align}
\label{eq:Hl}
\Ham_{\scl}(\Omega_1) = & \Ham_{\ff} +  \hbar  c\Omega_1  (b_V^{\dagger} b_V+ b_H^{\dagger} b_H  ) \non \\
 & +  i \hbar c  \Gamma_H^{1/2} \left( v_{\pos} b_V^{\dagger}  - v_{\pos}^{\dagger} b_V \right)  \non \\
& + i \hbar c \Gamma_V^{1/2} \left( h_{\pos} b_H^{\dagger} - h_{\pos}^{\dagger} b_H \right).
\end{align}
This Hamiltonian, which we have explicitly parametrized by the cavity frequency $\Omega_1$ for later convenience, is linear in the sense that the equations of motion which it generates for the field operators $h_z(\tau)$ and $v_z(\tau)$ are linear differential equations. Application of the corresponding inverse scattering matrix $\scatteringmatrixld(\Omega_1) \equiv \lim_{\tau \rightarrow \infty} e^{-i \Ham_{\ff}(\Omega_1) \tau / \hbar c } e^{+i \Ham_{\scl}(\Omega_1) \tau / \hbar c }$ then removes linear evolution from $\scatteringmatrixnl$.

\begin{figure}[tb]
\begin{center}
\subfigure[]{\label{fig:TimeReversedEvolution}
\includegraphics[scale=0.2]{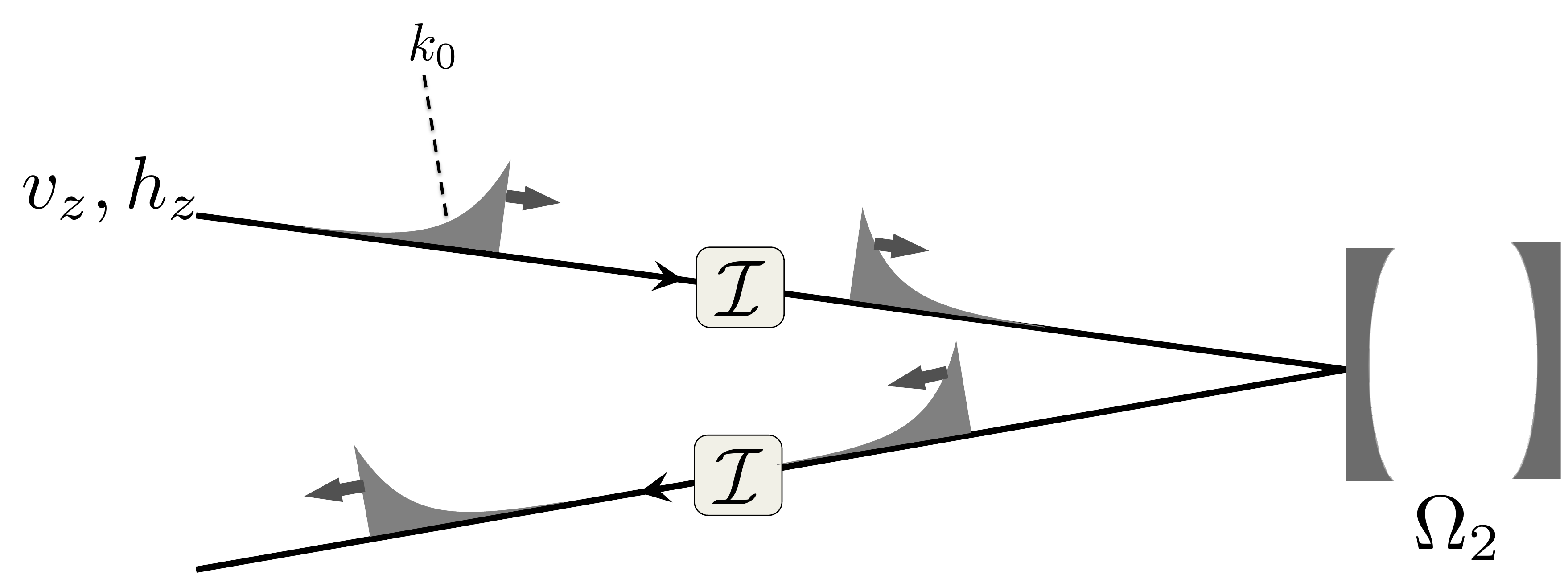} }  \\
\subfigure[]{\label{fig:pulseinverter}
\includegraphics[scale=.35]{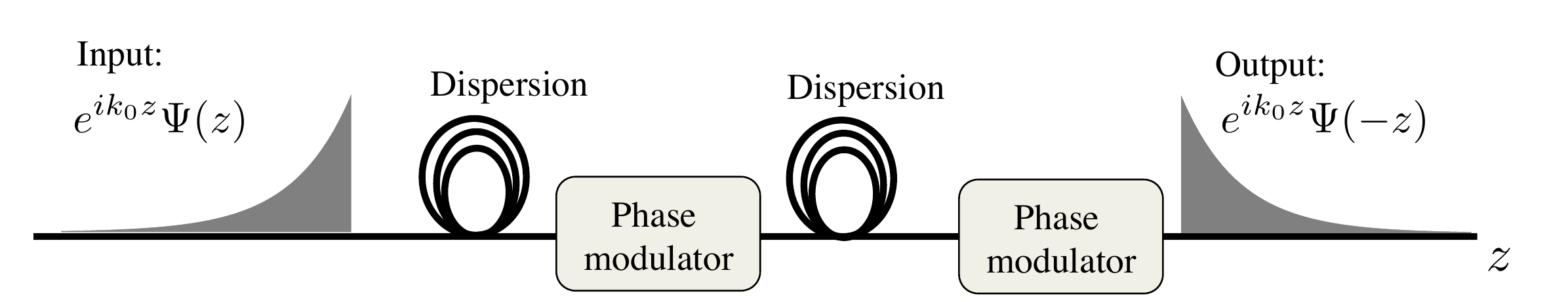} }  \\
\subfigure[]{\label{fig:spatialanalog}
\includegraphics[scale=.4]{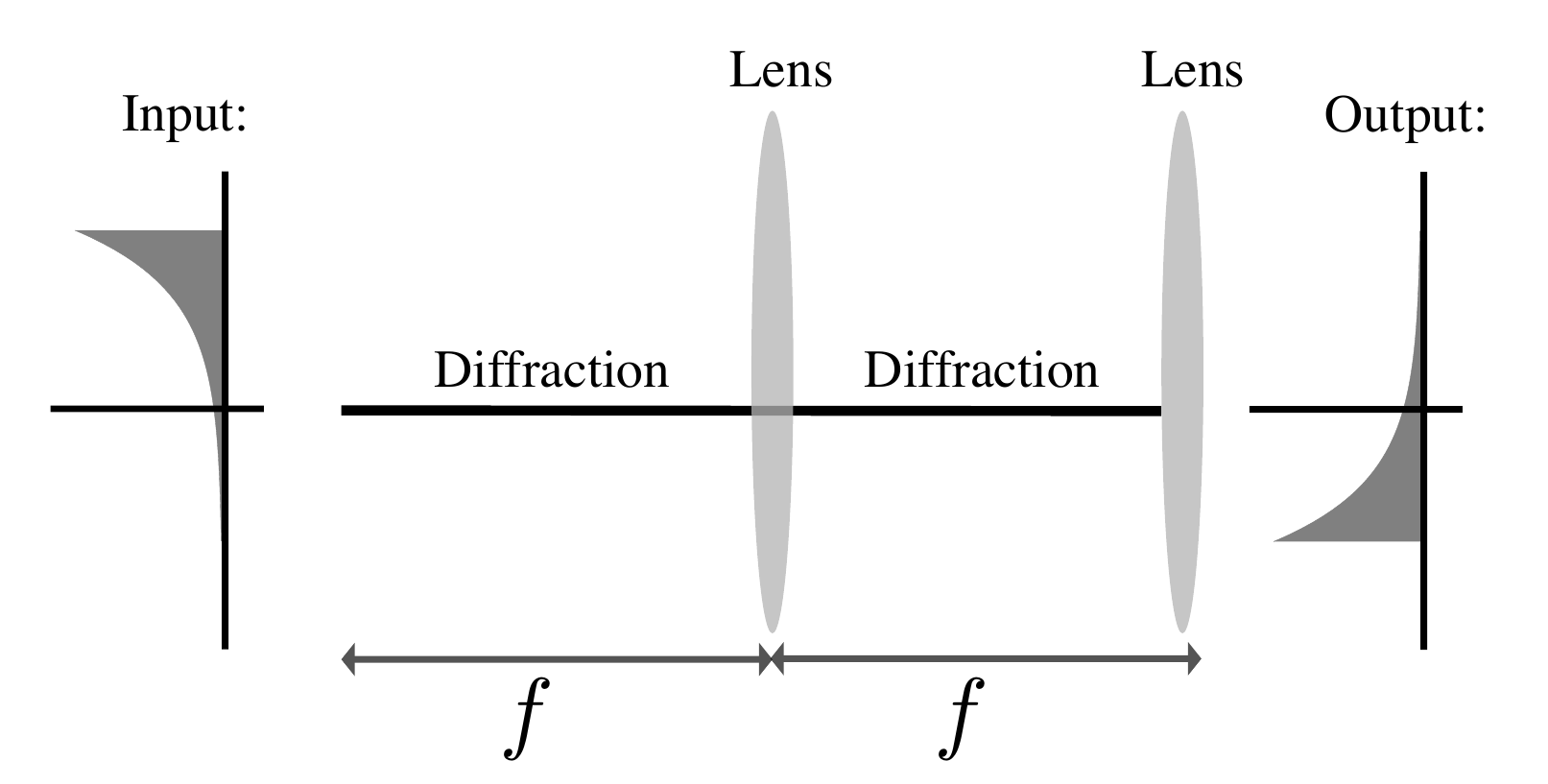}}
\end{center}
\caption{(a): Optical circuit to simulate time-reversed evolution under \Eqref{eq:Hl}. The cavity frequency $\Omega_2$ is chosen so that $k_0-\Omega_1 = -(k_0 - \Omega_2)$. (b): Temporal imaging system to realize inversion $\cI$ about center frequency $k_0$, using two dispersive delay lines and two quadratic phase modulators. (c): The analogous spatial imaging system using free-space diffraction and thin lenses.}
\end{figure}

This useful form of error-correction can, in principle, be implemented using linear optics. \Figureref{fig:TimeReversedEvolution} shows an optical circuit that removes linear evolution from input photons with center wavenumber $k_0$ by simulating time reversed evolution under \Eqref{eq:Hl}.  First, the baseband modulation of the input photon pulses is inverted ($\cI$). The pulses then interact with empty one-sided cavities, and, finally, the baseband modulation is re-inverted.

Real-space inversion of an optical pulse's baseband modulation corresponds to inversion  about its center wavenumber $k_0$ in Fourier-space. This transformation, 
\begin{align}
\label{eq:I}
\cI^{\dagger} \wt{h}_k \cI = \wt{h}_{2 k_0 - k},
\end{align}
can be achieved using temporal imaging \cite{Kolner1989,Kolner1994,Kauffman1994,Kuzucu2009}. Temporal imaging is the longitudinal analog of traditional spatial imaging: in spatial imaging, a beam's transverse profile is manipulated using free-space diffraction and thin lenses;  in temporal imaging, the longitudinal (temporal) profile is manipulated using dispersive delay lines and quadratic phase modulation. \Figureref{fig:pulseinverter} shows a temporal imaging system for baseband modulation inversion, while \Figref{fig:spatialanalog} shows its spatial analog. 

While this method has not, to our knowledge, been used to demonstrate pulse inversion with quantum light, we see no fundamental physical principle preventing its implementation. Because the scheme to implement $\cI$ shown in \Figref{fig:pulseinverter} involves only passive linear field transformations (dispersion and phase modulation) it behaves identically with respect to classical fields and few-photon pulses. 

%While optical pulse inversion has not, to our knowledge, been demonstrated experimentally with single photon pulses, we see no fundamental physical principle preventing its implementation.Because $\cI$ can be realized using only linear field transformations (dispersion and phase modulation) noise associated with nonlinear quantum processes \cite{Boivin1994,JHS2006,needmore} should be avoidable.
%%%The last sentence is horribly vague.

After inverting the optical pulses, the fields in \Figref{fig:TimeReversedEvolution} evolve forward in time under the linearized Hamiltonian $\Ham_{\scl}(\Omega_2)$ with cavity frequency $\Omega_2$. This corresponds to applying   $\scatteringmatrix_{\scl}(\Omega_2)$ on the field operators. Because the  equations of motion generated by $\Ham_{\scl}(\Omega_2)$ are linear, the mapping of the \emph{field operators} under $\scatteringmatrix_{\scl}(\Omega_2)$ is analogous to the mapping of single photon  packets under $\scatteringmatrix_{\scnl}$ (\Eqref{eq:kspaceprop}):
\begin{align}
\label{eq:Sl}
\scatteringmatrix_{\scl}^{\dagger}(\Omega_2) \wt{h}_k  \scatteringmatrix_{\scl}(\Omega_2)  = \wt{h}_k \frac{k - (\Omega_2 + i \Gamma_H/2)}{k - (\Omega_2 - i \Gamma_H/2)},
\end{align}
and similarly for $v_k$. By picking the pulse center wavenumber $k_0$, atomic resonance $\Omega_1$, and cavity resonance $\Omega_2$ such that photon-atom and photon-cavity detunings are equal and opposite, viz. $k_0-\Omega_1 = -(k_0 - \Omega_2)$, the combined effect of pulse inversion, followed by evolution under $\Ham_{\scl}(\Omega_2)$, followed by pulse inversion yields time-reversed evolution under $\Ham_{\scl}(\Omega_1)$:
\begin{align}
\label{eq:timereversal}
\cI \scatteringmatrix_{\scl}(\Omega_2) \cI = \scatteringmatrix_{\scl}^{\dagger}(\Omega_1).
\end{align}
In this way, the linear portion of $\scatteringmatrixnl$ can be undone.

\subsection{The Primitive Gate}
The combined effect of nonlinear interaction with the $\vee$-system  and removal of linear evolution is evolution under $\scatteringmatrixld \scatteringmatrixnl$:
\begin{subequations}
\label{eq:gatefield}
\begin{align}
\ket{\vac}\xrightarrow{\scatteringmatrixld \scatteringmatrixnl } & \ket{\vac} \\
\ket{H}\xrightarrow{\scatteringmatrixld \scatteringmatrixnl } & \ket{H} \\
\ket{V}\xrightarrow{\scatteringmatrixld \scatteringmatrixnl } & \ket{V} \\
\ket{HV}\xrightarrow{\scatteringmatrixld \scatteringmatrixnl } &(1-|\veps|^2) e^{i \phi_{\scnl} } \ket{HV} + \veps \ket{e}.
\end{align}
\end{subequations}
Here $\ket{e}$ is a two-photon state whose presence reflects errors intrinsic to the nonlinear evolution only. We refer to the transformation \Eqref{eq:gatefield} as our \emph{primitive} conditional $\phi_{\scnl}$-phase gate; this gate is primitive in the sense that it is not built by cascading smaller gates.

It is convenient to describe the primitive gate as transformation on the logical subspace $\{ \ket{\vac}, \ket{H}, \ket{V}, \ket{HV} \}$ alone. For nonzero errors $\veps$,  the mapping \Eqref{eq:gatefield} between input and output field states is not unitary when restricted to the this subspace,  because of pulse deformation and undesirable entanglement generated between continuous degrees of freedom (e.g., photon momentum) When restricted to the logical subspace, \Eqref{eq:gatefield} corresponds to a trace-preserving quantum operation $\cE_{\tprim}$:
\begin{align}
\cE_{\tprim} (\rho) = U_{\phi_{\scnl}} \left( E_1 \rho E_1^{\dagger}  +  E_2 \rho E_2^{\dagger} \right) U_{\phi_{\scnl}}^{\dagger}.
\end{align}
Here $\rho$ is a two-qubit density matrix, $U_{\phi}$ is the ideal conditional $\phi$-phase gate, and the operation elements $\{E_1,E_2\}$ represent pure amplitude damping of the two-photon state $\ket{HV}$ out of the logical subspace. In the usual basis,
\begin{subequations}
\begin{align}
U_{\phi} = &
\begin{bmatrix}
1 & 0 & 0 & 0 \\ 
0 & 1 & 0 & 0 \\
0 & 0 & 1 & 0 \\
0 & 0 & 0 & e^{i \phi} 
\end{bmatrix}, \\
E_1  = &
\begin{bmatrix}
1 & 0 & 0 & 0 \\ 
0 & 1 & 0 & 0 \\
0 & 0 & 1 & 0 \\
0 & 0 & 0 & (1-|\veps|^2)^{1/2}
\end{bmatrix} \\
E_2  =  &
\begin{bmatrix}
0 & 0 & 0 & \veps \\ 
0 & 0 & 0 & 0 \\
0 & 0 & 0 & 0 \\
0 & 0 & 0 & 0
\end{bmatrix}.
\end{align}
\end{subequations}
This operator-sum representation of the primitive gate is useful in determining its fidelity with an ideal conditional phase gate.

\subsection{Fidelity of a Single Gate}
\label{sec:gatefid}
The \emph{fidelity} of two states is a measure of how close they are to one another, increasing from 0 (orthogonal states) to 1 (identical states). The fidelity of a pure state $\psi$ with a mixed state $\rho$ may be defined as their overlap, $F(\ket{\psi}, \rho) = \bra{\psi} \rho \ket{\psi}$. \emph{Gate fidelity} extends this idea from states to logical operations on qubits. The (minimum) gate fidelity of a quantum operation $\cE$ with a unitary gate $U$ that $\cE$ approximates is the fidelity of $\cE$'s output with the target output, minimized over pure state inputs \cite{NC}:
\begin{align}
F( \cE, U) = \min_{\ket{\psi}} \bra{ \psi} U^{\dagger} \cE(\proj{\psi} ) U \ket{\psi}.
\end{align}
The infidelity $1-F( \cE, U)$ is  the (maximum) probability that the $\cE$ fails to effect the desired transformation $U$.

 The fidelity of our gate $\cE_{\tprim}$ with the ideal conditional phase gate $U_{\phi_{\scnl}}$ is
\begin{align}
F(\cE_{\tprim} ,U_{\phi_{\scnl}}) \equiv & \min_{ \ket{\psi} } \bra{\psi} U_{\phi_{\scnl}}^{\dagger} \cE_{\tprim} (\proj{\psi}) U_{\phi_{\scnl}} \ket{\psi} \non \\
= & \min_{\ket{\psi}} \left[ |\bra{ \psi} E_1 \ket{\psi}|^2 + |\bra{ \psi} E_2 \ket{\psi}|^2 \right] \non \\
= & 1 - |\veps|^2.
\end{align}
Here the minimizing state is $\ket{11}_{\scl}=\ket{HV}$.

We now consider the relationship between the fidelity $F(\cE_{\tprim} ,U_{\phi_{\scnl}})$ and the nonlinear phase shift when the real-space principal mode  wavefunction $\psi(z)$ in \Eqref{eq:principalmode} is a rising exponential with center wavenumber $k_0$ and width $\gamma$:
\begin{align}
\label{eq:principalmode}
\psi(z) \equiv e^{i k_0 z} \Psi(z), && \Psi(z) = \theta(-z) e^{-\gamma | z|/2}.
\end{align}
This particular principal mode wavefunction is chosen because, as demonstrated in the next section, it is possible to make a projective measurement that distinguishes excitations of this principal mode from all other modes. Additionally, we now specialize to the case in which $\Gamma_H = \Gamma_V \equiv \Gamma$ in order that the qubits are treated symmetrically.

\begin{figure}[tb]
\begin{center}
\subfigure[]{\label{fig:fidplot1}
\includegraphics[scale=0.66]{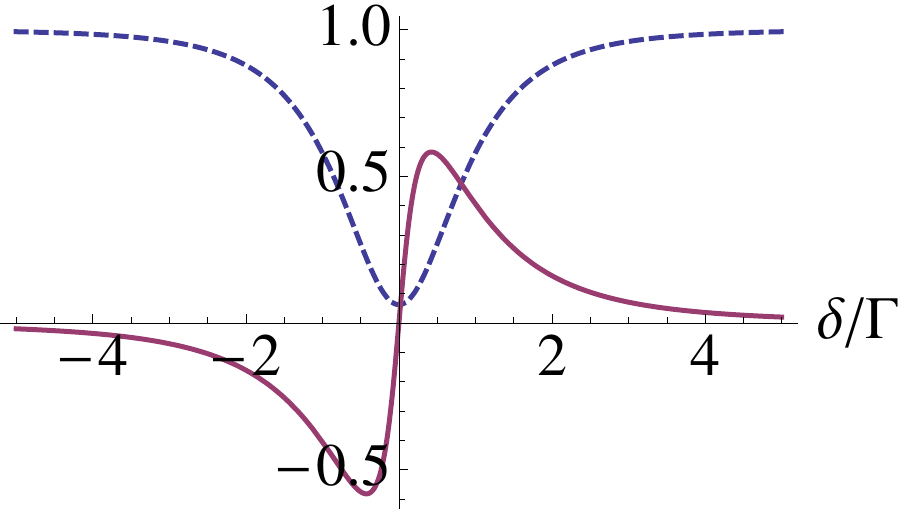} }
\subfigure[]{\label{fig:purityplot1}
\includegraphics[scale=0.66]{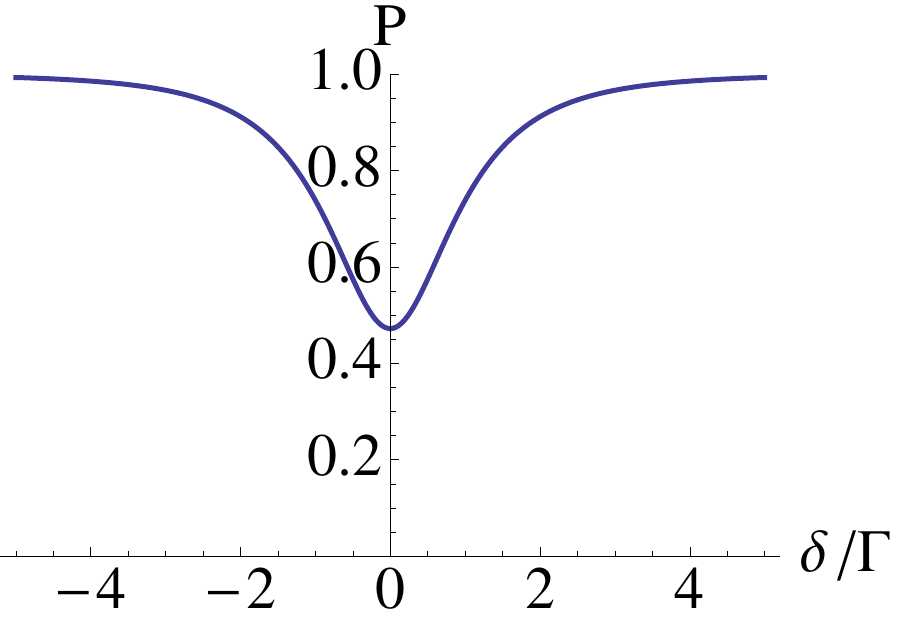} }
\caption{(a): Comparison of the fidelity (dashed) and nonlinear phase (solid) when $\gamma = \Gamma$ as a function of the detuning $\delta$. (b): Purity $P = \tr \rho_H'^2$ of the horizontal photon's output density matrix when the input state is $\ket{HV}$ as a function of $\delta$ when $\gamma=\Gamma$.}
\end{center}
\end{figure}

\textbf{Large Phase Shifts.}
If the fidelity $F(\cE_{\tprim} ,U_{\phi_{\scnl}})$ and phase shift $\phi_{\scnl}$ could both be large simultaneously, the primitive gate would be an effective conditional phase gate.

It is only when the atomic line-width $\Gamma$ is comparable in size to the pulse bandwidth $\gamma$ that a large nonlinear phase shift is possible. If $\gamma \gg \Gamma$, then the pulse is too broadband to interact significantly with the atom, while if $\gamma \ll \Gamma$, then one sees from \Eqref{eq:twophotonpropagator} that the range $\Gamma^{-1}$ of the nonlinear piece of the two-photon propagator is negligible in comparison to the pulse length $\gamma^{-1}$.

 \Figureref{fig:fidplot1} shows the fidelity and nonlinear phase shift as  functions of the detuning $\delta \equiv k_0 - \Omega_1$ in the particular case $\gamma = \Gamma$. The phase shift $\phi_{\scnl}(\delta)$ has the form of a dispersion curve, while the infidelity $1-F$ mimics an absorption curve. The figure shows that while large nonlinear phase shifts are possible for nearly resonant pulses, the fidelity is unacceptably low in these cases---a conclusion similar to those drawn in \cite{JHS2006,JGB2010}. 
 
 A large contribution to this fidelity degradation is the entanglement generated between the position (or momentum) coordinates of the horizontally and vertically polarized photons. This entanglement reflects antibunching in the two-photon output wavefunction, and is characterized by a sub-unity purity $P \equiv \tr {\rho'}_{H}^2$ of the horizontal photon's output density matrix $\rho'_H \equiv \tr_V \cE( \proj{HV} )$  (\Figref{fig:purityplot1}).

\textbf{Small Phase Shifts.}
While large phase shifts are accompanied by large errors, it is possible to achieve small phase shifts with a much smaller error: $ |\veps|^2\ll \phi_{\scnl}  \ll 1$. When the phase shift and error are small, it is convenient to write
\begin{align}
\bra{HV} \scatteringmatrixld \scatteringmatrixnl \ket{HV} = 1 + i \zeta,
\end{align}
so that to lowest order in $\zeta$ the nonlinear  phase shift is $\phi_{\scnl} =  \re[\zeta]$ and error probability $1-F$ is $|\veps|^2 =   2 \im{[\zeta]}$. 

Particularly simple expressions for the phase shift and error are obtained when the pulse bandwidth is much less than the atomic line-width, $\gamma \ll \Gamma$. Because the photon wavefunction $\psi(z)$ has length $\sim \gamma^{-1}$ and is normalized to unity, this can be considered a sort of weak-excitation regime. In this case, $\zeta$ is readily calculated from the Fourier-space propagators, \Eqref{eq:kspaceprop} and \Eqref{eq:kspace2photon}.  One finds that in this case dependences of $\phi_{\scnl}$ and $|\veps|^2$ on the detuning are again those of dispersion and absorption curves:
\begin{subequations}
\label{eq:weakexcitation}
\begin{align}
\phi_{\scnl} = &  \frac{\gamma \Gamma^2 \delta }{[\delta^2+(\Gamma/2)^2]^2}, \\
|\veps|^2 = & \frac{\Gamma}{\delta} \phi_{\scnl},
\end{align}
\end{subequations}
to lowest nonvanishing order in $\gamma/\Gamma$ .  

From \Eqref{eq:weakexcitation} it is clear that when $\gamma \ll \Gamma \ll \delta$ the nonlinear phase shift, while very small, is large in comparison to the error probability: $|\veps|^2 \ll \phi_{\scnl}$. Actually, the relation $|\veps|^2 \ll \phi_{\scnl}$ can be achieved without requiring that $\gamma \ll \Gamma$: it is enough for the photons to be far-detuned. When $\Gamma, \gamma \ll \delta$, we have
\begin{subequations}
\label{eq:overlaps}
\begin{align}
\phi_{\scnl} = & \re[\zeta] = \frac{\gamma \Gamma^2}{\delta^3} \left(\frac{1+5\frac{\gamma}{\Gamma} }{1+\frac{\gamma}{\Gamma}} \right), \\
|\veps|^2 = &  2 \im{[\zeta]} = \frac{\Gamma}{\delta} \left( \frac{1+10 \frac{\gamma}{\Gamma}+\frac{\gamma^2}{\Gamma^2} }{1+5\frac{\gamma}{\Gamma}}\right) \phi_{\scnl},
\end{align}
\end{subequations}
to lowest order in $\max{[\gamma,\Gamma]}/\delta$. Again the nonlinear phase shift, though small, is much larger than the infidelity $|\veps|^2$. In this sense, our primitive conditional phase gate can be considered high-fidelity for small phase shifts.

\section{Cascading Small Phase Shifts}
The error $|\veps|^2$ in the primitive conditional phase gate discussed above is the probability that the gate causes the  two-photon state $\ket{HV}$ to leak out of the principal mode subspace. Because this error probability can be made much smaller than the phase shift in the far-detuned regime, the possibility  of cascading  $N=\pi/\phi_{\scnl}$ primitive gates to produce a high-fidelity conditional $\pi$-phase gate arises.

 When the primitive gate $\scatteringmatrixld \scatteringmatrixnl$ is cascaded $N$ times two sorts of errors  can occur. With each application, the probability of photons leaking out of the principal mode subspace increases; for small $|\veps|^2$, these \emph{leakage errors}  grow as $N |\veps|^2 = \pi |\veps|^2/\phi_{\scnl}  \ll 1$, and are not terribly problematic. However, amplitude that leaked from the principal mode subspace in earlier applications of $\scatteringmatrixld \scatteringmatrixnl$ can return in later applications with corrupted phase; these \emph{coherent feedback errors} can grow as $N^2 |\veps|^2$, which is not small. Alternatively, this difficulty can be seen by noting that the primitive gate cascaded $N$ times does \emph{not} correspond to  the quantum operation $\cE_{\tprim}$ cascaded $N$ times. This is because the state of the auxiliary modes changes with each application of the $\scatteringmatrixld \scatteringmatrixnl$.

\subsection{A Cascadable Primitive Gate}

We propose to eliminate coherent feedback errors by measuring the number of photons present in the auxiliary modes after each application of the primitive gate. For the sake of the following analysis, the result of this measurement need not be considered, only that with probability at least $1-|\veps|^2$ it projects the quantum state back onto the principal mode subspace. For this reason, we call this measurement process \emph{principal mode projection} (PMP). Performing PMP after each application of the primitive gate is a sort of Zeno effect error correction that prevents amplitude from leaking out of the principal mode subspace too quickly.

Crucially, the measurement used to implement PMP must be done in such a way that it is insensitive to the number of photons in the principal modes. If the principal mode function $\psi(z)$ is chosen to be the one-sided exponential used above (\Eqref{eq:principalmode}), then such a measurement can, in fact, be performed using empty optical cavities, the pulse inverter $\cI$ introduced in \Secref{sec:removinglinearevolution}, and irises.  

\begin{figure}[t]
\begin{center}
\includegraphics[scale=0.22]{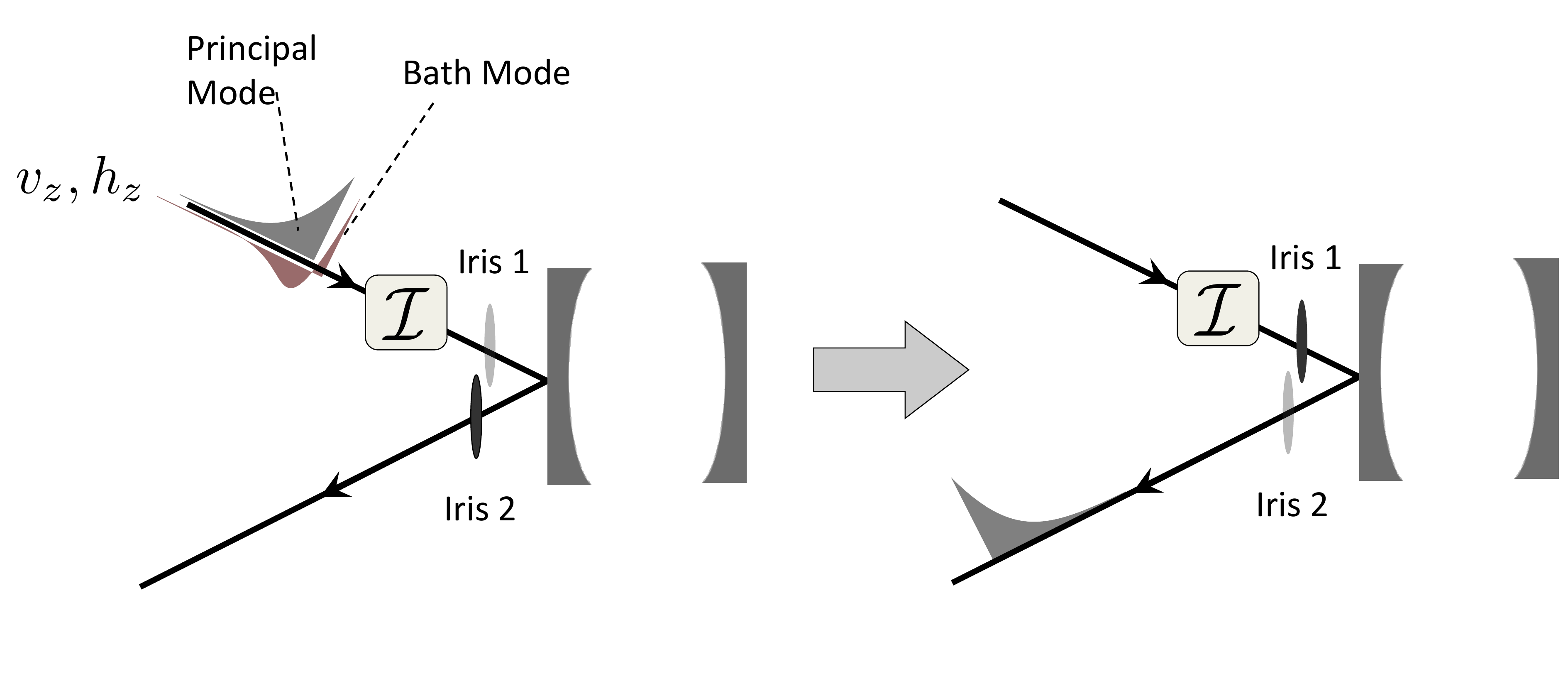}
\caption{Schematic of a principal mode projector  for  mode function given by  \Eqref{eq:principalmode}. Initially, iris 1 is open, and photons from the principal mode are absorbed by the cavity after inversion. After principal mode photons have been absorbed, iris 1 is shut and iris 2 is open, allowing cavity photons to be re-emitted into the principal mode.}
\label{fig:PMP}
\end{center}
\end{figure}

The scheme, illustrated in \Figref{fig:PMP}, exploits the fact that (ignoring free-space evolution) a cavity with resonant wavenumber $k_0$ and decay rate $\gamma$ preferentially emits photons with mode functions $e^{i k_0 z} \Psi(z)$ and preferentially absorbs from the inverted mode $e^{i k_0 z} \Psi(-z)$. This selectivity is used to load all principal mode photons into optical cavities. Once this is done, iris 1 is closed, preventing  non-principal mode photons from entering the cavity, while iris 2 is opened, allowing the cavity photons to be emitted back into the principal modes.  This setup could be modified to record the result of the PMP measurement, allowing for heralded operation and post-selection. However, the point of the present analysis is to provide a design for a deterministic gate, and thus our process employs no post-selection.
 
 \begin{figure}[tb]
\begin{center}
\includegraphics[scale=0.36]{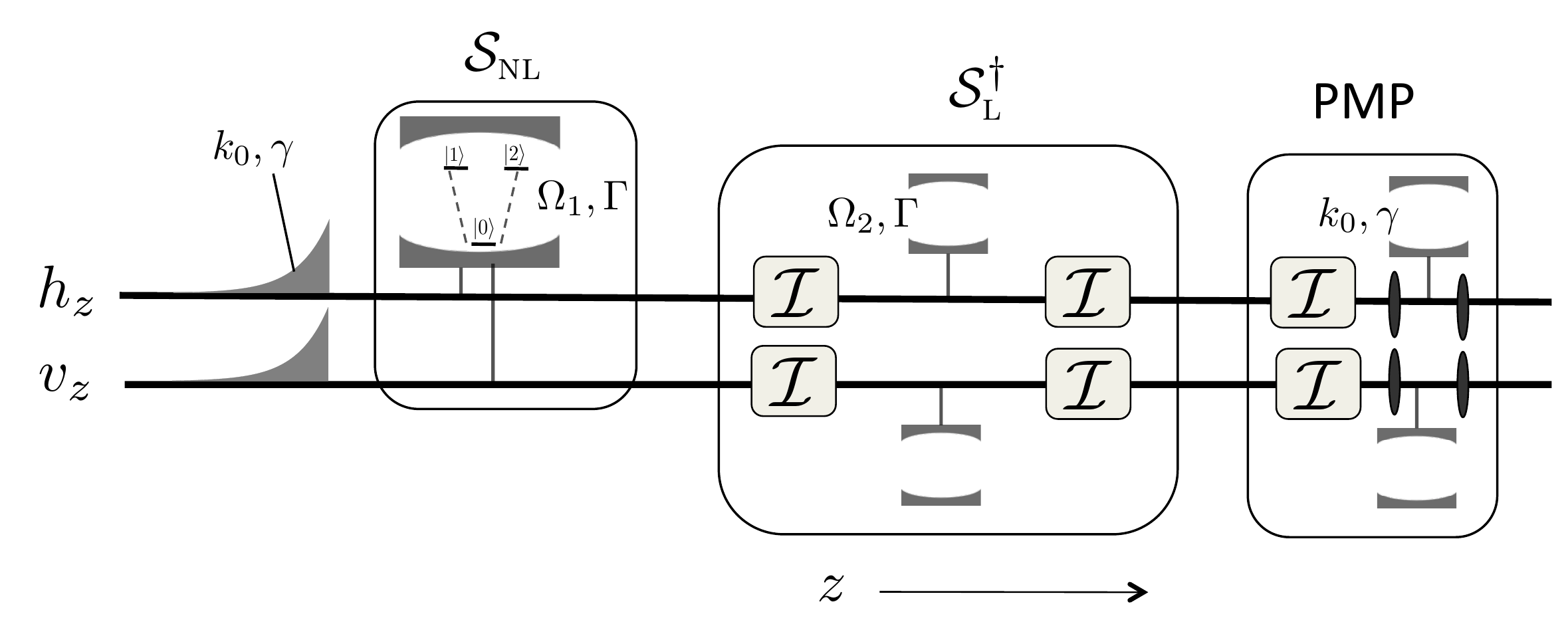}
\caption{The cascadable primitive gate. First, nonlinear evolution is provided by interaction with the atomic $\vee$-system. Linear evolution is then removed. Finally, principal mode projection is performed.}
\label{fig:wholegate}
\end{center}
\end{figure}

\Figureref{fig:wholegate} shows the entire process: interaction with the $\vee$-system ($\scatteringmatrixnl$), followed by removal of linear evolution ($\scatteringmatrixld$), followed by PMP. (Note that the second and third pulse inverters cancel, and thus need not actually be implemented.) This gate, which we call the \emph{cascadable primitive} gate is most naturally represented by a non-trace-preserving quantum operation \cite{NC},
\begin{align}
\cE_{\tcascprim}(\rho) = U_{\phi_{\scnl}}  E_1 \rho E_1^{\dagger} U_{\phi_{\scnl}}^{\dagger}.
\end{align}
where $\tr{[\cE_{\tcascprim}(\rho)]}$ is the probability  of success, i.e., that the output state has been collapsed into the principal mode subspace. 

\subsection{Fidelity of the Cascaded Gate}

Because of the PMP, the cascadable primitive gate can be cascaded $N =\pi/\phi_{\scnl}$ times to produce a high-fidelity conditional $\pi$-phase gate.  Without any post-selection, the fidelity of this cascaded gate with the ideal conditional $\pi$-phase gate is  the probability that PMP success occurs $N$ times:
\begin{align}
\label{eq:pifid}
F(\cE_{\tcascprim}^N,U_{\pi}) = & F(\cE_{\tcascprim},U_{\phi_{\scnl}})^N \non \\
= & 1 - \pi  \frac{\Gamma}{\delta} \left( \frac{1+10 \frac{\gamma}{\Gamma}+\frac{\gamma^2}{\Gamma^2} }{1+5\frac{\gamma}{\Gamma}}\right),
\end{align}
to lowest nonvanishing order in $\max{[\gamma,\Gamma]}/\delta$. In the far-detuned regime, $\gamma,\Gamma \gg \delta$, this fidelity can become quite large: cascading $\cE_{\tcascprim}$ can yield a high-fidelity conditional $\pi$-phase gate.

 Unfortunately, because the $\vee$-system's nonlinearity is so weak, an incredible number of cascades are required to produce a high fidelity conditional $\pi$-phase gate. For fixed $N$, \Eqref{eq:pifid} can be rewritten, after optimizing the ratio $\gamma/\Gamma$, as 
\begin{align}
\label{eq:impractical}
F(\cE_{\tcascprim}^N,U_{\pi}) \approx 1- (4.82) N^{-1/3}.
\end{align}
To achieve a fidelity greater than 95\% over $10^6$ cascades are required. The origin of this unfortunate scaling is the weak cross phase shift, $\phi_{\scnl} \propto \delta^{-3}$. If instead the phase shift and error were $\phi_{\scnl} \propto \delta^{-m}$ and $|\veps|^2 \propto \delta^{-n}$, the fidelity of the cascaded gate would be $F \sim 1 - N^{1-n/m}$.

Our cascadable primitive gate $\cE_{\tcascprim}$ operates in the far-detuned regime and incorporates two error correcting steps: the removal of linear evolution ($\scatteringmatrixld$) and the PMP. Principal mode projection is absolutely essential in making this gate cascadable. How important is removing the linear evolution? For the mode function used above, the linear errors $\{|\veps_H|^2,|\veps_V|^2\}$ must be removed. Because the Fourier-space mode function $\wt{\psi}(k) = i\gamma^{1/2} (k-k_0+i \gamma/2)^{-1}$ falls off only as $k^{-1}$, linear errors are of the same order of magnitude as the nonlinear phase shift. However, for more well behaved Fourier-space mode functions (e.g. Gaussians $\wt{\psi}(k) \sim \exp{[- (k-k_0)^2/4\gamma^2 }]$ and even Lorentzians $\wt{\psi}(k) \sim [(k-k_0)^2+\gamma^2]^{-1}$) linear errors are of the same order as nonlinear errors. If PMPs could be constructed for these modes, the removal of linear evolution would not be essential.

\section{Conclusions}
Treating light as a multimode quantum field, we have described conditional phase gates in which  photonic qubits interact with a three-level $\vee$-system. Although we have used the language of atomic and optical systems in our analysis, other implementations are possible.  In the microwave, for example, the one-dimensional field of transmission line waveguides have been coupled to artificial atoms \cite{Shen2005, Nakamura2010}.

 In the regime of large nonlinear phase shifts, our primitive (non-cascaded) gate  has unacceptably low fidelity, as has been found for other gates relying on quantum cross-phase modulation \cite{JHS2006,JHS2007,JGB2010}. We attribute much of this infidelity to undesirable  entanglement generated by the local character of the  nonlinear interaction between the horizontal and vertically polarized fields. 

In contrast, the primitive gate can produce a small nonlinear phase shift with very high fidelity ($1-F \ll \phi_{\scnl}$) by operating in the far-detuned regime.  However, one cannot straightforwardly cascade this high-fidelity, small conditional phase shift because of  coherent feedback errors that grow as $N^2$.

We have shown that it is, in principle, possible to overcome the cascadability problem by making a projective measurement of the bath modes' photon number after each small conditional phase gate. With high probability, this measurement projects the field state back onto the information-carrying principal modes. This step---principal mode projection---uses the quantum Zeno effect to prevent coherent feedback errors from occurring, making a cascadable primitive conditional phase gate.

% Unfortunately, while this gate is cascadable in theory, the $\vee$-system's weak nonlinearity may make cascading our gate impractical: for a fixed number $N$ of cascades, the cascaded gate's fidelity is $F \sim 1-N^{-1/3}$.

We suggest that principal mode projection could be a helpful subroutine in the future of photonic quantum information processing. While the $\vee$-system's weak cross-phase shift may make cascading our gate impractical (\Eqref{eq:impractical}), PMP together with stronger nonlinearities, e.g., the giant Kerr effect,  could potentially realize a conditional $\pi$-phase gate whose fidelity scales more favorably with $N$. Other interesting future directions include the possibility of considering alternative  PMP constructions and analyzing the usefulness of PMP in overcoming XPM noise in optical fiber \cite{JHS2006}. 
%Other interesting questions arise. Our PMP realization required a specific choice of the principal mode function. What other PMPs can be constructed? Could Zeno-effect error correction through PMP aid in overcoming XPM noise \cite{JHS2006} in optical fiber?  

\begin{acknowledgements}
This research was supported by the NSF IGERT program Interdisciplinary Quantum Information Science  and Engineering (iQuISE) and the DARPA Quantum Entanglement and Information Science Technology (QUEST) program.
\end{acknowledgements}

\bibliography{bibliography}

%Merlin.mbs v4.21 2009-07-09.
\begin{thebibliography}{10}%
\makeatletter
\providecommand \@ifxundefined [1]{%
 \ifx #1\undefined \expandafter \@firstoftwo
 \else \expandafter \@secondoftwo
\fi
}%
\providecommand \@ifnum [1]{%
 \ifnum #1\expandafter \@firstoftwo
 \else \expandafter \@secondoftwo
\fi
}%
\providecommand \enquote [1]{``#1''}%
\providecommand \bibnamefont  [1]{#1}%
\providecommand \bibfnamefont [1]{#1}%
\providecommand \citenamefont [1]{#1}%
\providecommand\href[0]{\@sanitize\@href}%
\providecommand\@href[1]{\endgroup\@@startlink{#1}\endgroup\@@href}%
\providecommand\@@href[1]{#1\@@endlink}%
\providecommand \@sanitize [0]{\begingroup\catcode`\&12\catcode`\#12\relax}%
\@ifxundefined \pdfoutput {\@firstoftwo}{%
 \@ifnum{\z@=\pdfoutput}{\@firstoftwo}{\@secondoftwo}%
}{%
 \providecommand\@@startlink[1]{\leavevmode\special{html:<a href="#1">}}%
 \providecommand\@@endlink[0]{\special{html:</a>}}%
}{%
 \providecommand\@@startlink[1]{%
  \leavevmode
  \pdfstartlink
   attr{/Border[0 0 1 ]/H/I/C[0 1 1]}%
   user{/Subtype/Link/A<</Type/Action/S/URI/URI(#1)>>}%
  \relax
 }%
 \providecommand\@@endlink[0]{\pdfendlink}%
}%
\providecommand \url  [0]{\begingroup\@sanitize \@url }%
\providecommand \@url [1]{\endgroup\@href {#1}{\urlprefix}}%
\providecommand \urlprefix [0]{URL }%
\providecommand \Eprint[0]{\href }%
\@ifxundefined \urlstyle {%
  \providecommand \doi [1]{doi:\discretionary{}{}{}#1}%
}{%
  \providecommand \doi [0]{doi:\discretionary{}{}{}\begingroup
  \urlstyle{rm}\Url }%
}%
\providecommand \doibase [0]{http://dx.doi.org/}%
\providecommand \Doi[1]{\href{\doibase#1}}%
\providecommand \bibAnnote [3]{%
  \BibitemShut{#1}%
  \begin{quotation}\noindent
    \textsc{Key:}\ #2\\\textsc{Annotation:}\ #3%
  \end{quotation}%
}%
\providecommand \bibAnnoteFile [2]{%
  \IfFileExists{#2}{\bibAnnote {#1} {#2} {\input{#2}}}{}%
}%
\providecommand \typeout [0]{\immediate \write \m@ne }%
\providecommand \selectlanguage [0]{\@gobble}%
\providecommand \bibinfo [0]{\@secondoftwo}%
\providecommand \bibfield [0]{\@secondoftwo}%
\providecommand \translation [1]{[#1]}%
\providecommand \BibitemOpen[0]{}%
\providecommand \bibitemStop [0]{}%
\providecommand \bibitemNoStop [0]{.\EOS\space}%
\providecommand \EOS [0]{\spacefactor3000\relax}%
\providecommand \BibitemShut [1]{\csname bibitem#1\endcsname}%
%</preamble>
\bibitem{GJMilburn1989}%
  \BibitemOpen
  \bibfield{author}{%
  \bibinfo {author} {\bibfnamefont{G.~J.}\ \bibnamefont{Milburn}},\ }%
  \bibfield{journal}{%
  \Doi{10.1103/PhysRevLett.62.2124}{\bibinfo {journal} {Phys. Rev. Lett.}}\ }%
  \textbf{\bibinfo {volume} {62}},\ \bibinfo {pages} {2124} (\bibinfo {year}
  {1989})%
  \bibAnnoteFile{NoStop}{GJMilburn1989}%
\bibitem{Turchette1995}%
  \BibitemOpen
  \bibfield{author}{%
  \bibinfo {author} {\bibfnamefont{Q.~A.}\ \bibnamefont{Turchette}}, \bibinfo
  {author} {\bibfnamefont{C.~J.}\ \bibnamefont{Hood}}, \bibinfo {author}
  {\bibfnamefont{W.}~\bibnamefont{Lange}}, \bibinfo {author}
  {\bibfnamefont{H.}~\bibnamefont{Mabuchi}},\ and\ \bibinfo {author}
  {\bibfnamefont{H.~J.}\ \bibnamefont{Kimble}},\ }%
  \bibfield{journal}{%
  \Doi{10.1103/PhysRevLett.75.4710}{\bibinfo {journal} {Phys. Rev. Lett.}}\ }%
  \textbf{\bibinfo {volume} {75}},\ \bibinfo {pages} {4710} (\bibinfo {year}
  {1995})%
  \bibAnnoteFile{NoStop}{Turchette1995}%
\bibitem{Chuang1995}%
  \BibitemOpen
  \bibfield{author}{%
  \bibinfo {author} {\bibfnamefont{I.~L.}\ \bibnamefont{Chuang}}\ and\ \bibinfo
  {author} {\bibfnamefont{Y.}~\bibnamefont{Yamamoto}},\ }%
  \bibfield{journal}{%
  \Doi{10.1103/PhysRevA.52.3489}{\bibinfo {journal} {Phys. Rev. A}}\ }%
  \textbf{\bibinfo {volume} {52}},\ \bibinfo {pages} {3489} (\bibinfo {year}
  {1995})%
  \bibAnnoteFile{NoStop}{Chuang1995}%
\bibitem{Lukin2000}%
  \BibitemOpen
  \bibfield{author}{%
  \bibinfo {author} {\bibfnamefont{M.~D.}\ \bibnamefont{Lukin}}\ and\ \bibinfo
  {author} {\bibfnamefont{A.}~\bibnamefont{Imamo\ifmmode~\breve{g}\else
  \u{g}\fi{}lu}},\ }%
  \bibfield{journal}{%
  \Doi{10.1103/PhysRevLett.84.1419}{\bibinfo {journal} {Phys. Rev. Lett.}}\ }%
  \textbf{\bibinfo {volume} {84}},\ \bibinfo {pages} {1419} (\bibinfo {year}
  {2000})%
  \bibAnnoteFile{NoStop}{Lukin2000}%
\bibitem{Kimble2004}%
  \BibitemOpen
  \bibfield{author}{%
  \bibinfo {author} {\bibfnamefont{L.-M.}\ \bibnamefont{Duan}}\ and\ \bibinfo
  {author} {\bibfnamefont{H.~J.}\ \bibnamefont{Kimble}},\ }%
  \bibfield{journal}{%
  \Doi{10.1103/PhysRevLett.92.127902}{\bibinfo {journal} {Phys. Rev. Lett.}}\
  }%
  \textbf{\bibinfo {volume} {92}},\ \bibinfo {pages} {127902} (\bibinfo {year}
  {2004})%
  \bibAnnoteFile{NoStop}{Kimble2004}%
\bibitem{Koshino2010}%
  \BibitemOpen
  \bibfield{author}{%
  \bibinfo {author} {\bibfnamefont{K.}~\bibnamefont{Koshino}}, \bibinfo
  {author} {\bibfnamefont{S.}~\bibnamefont{Ishizaka}},\ and\ \bibinfo {author}
  {\bibfnamefont{Y.}~\bibnamefont{Nakamura}},\ }%
  \bibfield{journal}{%
  \Doi{10.1103/PhysRevA.82.010301}{\bibinfo {journal} {Phys. Rev. A}}\ }%
  \textbf{\bibinfo {volume} {82}},\ \bibinfo {pages} {010301} (\bibinfo {year}
  {2010})%
  \bibAnnoteFile{NoStop}{Koshino2010}%
\bibitem{JHS2006}%
  \BibitemOpen
  \bibfield{author}{%
  \bibinfo {author} {\bibfnamefont{J.~H.}\ \bibnamefont{Shapiro}},\ }%
  \bibfield{journal}{%
  \Doi{10.1103/PhysRevA.73.062305}{\bibinfo {journal} {Phys. Rev. A}}\ }%
  \textbf{\bibinfo {volume} {73}},\ \bibinfo {pages} {062305} (\bibinfo {year}
  {2006})%
  \bibAnnoteFile{NoStop}{JHS2006}%
\bibitem{JHS2007}%
  \BibitemOpen
  \bibfield{author}{%
  \bibinfo {author} {\bibfnamefont{J.~H.}\ \bibnamefont{Shapiro}}\ and\
  \bibinfo {author} {\bibfnamefont{M.}~\bibnamefont{Razavi}},\ }%
  \bibfield{journal}{%
  \bibinfo {journal} {New J. Phys.}\ }%
  \textbf{\bibinfo {volume} {9}},\ \bibinfo {pages} {16} (\bibinfo {year}
  {2007})%
  \bibAnnoteFile{NoStop}{JHS2007}%
\bibitem{JGB2010}%
  \BibitemOpen
  \bibfield{author}{%
  \bibinfo {author} {\bibfnamefont{J.}~\bibnamefont{Gea-Banacloche}},\ }%
  \bibfield{journal}{%
  \Doi{10.1103/PhysRevA.81.043823}{\bibinfo {journal} {Phys. Rev. A}}\ }%
  \textbf{\bibinfo {volume} {81}},\ \bibinfo {pages} {043823} (\bibinfo {year}
  {2010})%
  \bibAnnoteFile{NoStop}{JGB2010}%
\bibitem{Koshino2009}%
  \BibitemOpen
  \bibfield{author}{%
  \bibinfo {author} {\bibfnamefont{K.}~\bibnamefont{Koshino}},\ }%
  \bibfield{journal}{%
  \Doi{10.1103/PhysRevA.80.023813}{\bibinfo {journal} {Phys. Rev. A}}\ }%
  \textbf{\bibinfo {volume} {80}},\ \bibinfo {pages} {023813} (\bibinfo {year}
  {2009})%
  \bibAnnoteFile{NoStop}{Koshino2009}%
\bibitem{Schmidt1996}%
  \BibitemOpen
  \bibfield{author}{%
  \bibinfo {author} {\bibfnamefont{H.}~\bibnamefont{Schmidt}}\ and\ \bibinfo
  {author} {\bibfnamefont{A.}~\bibnamefont{Imamogdlu}},\ }%
  \bibfield{journal}{%
  \Doi{10.1364/OL.21.001936}{\bibinfo {journal} {Opt. Lett.}}\ }%
  \textbf{\bibinfo {volume} {21}},\ \bibinfo {pages} {1936} (\bibinfo {year}
  {1996})%
  \bibAnnoteFile{NoStop}{Schmidt1996}%
\bibitem{Fleisch2005}%
  \BibitemOpen
  \bibfield{author}{%
  \bibinfo {author} {\bibfnamefont{M.}~\bibnamefont{Fleischhauer}}, \bibinfo
  {author} {\bibfnamefont{A.}~\bibnamefont{Imamoglu}},\ and\ \bibinfo {author}
  {\bibfnamefont{J.~P.}\ \bibnamefont{Marangos}},\ }%
  \bibfield{journal}{%
  \Doi{10.1103/RevModPhys.77.633}{\bibinfo {journal} {Rev. Mod. Phys.}}\ }%
  \textbf{\bibinfo {volume} {77}},\ \bibinfo {pages} {633} (\bibinfo {year}
  {2005})%
  \bibAnnoteFile{NoStop}{Fleisch2005}%
\bibitem{Vaidman1996}%
  \BibitemOpen
  \bibfield{author}{%
  \bibinfo {author} {\bibfnamefont{L.}~\bibnamefont{Vaidman}}, \bibinfo
  {author} {\bibfnamefont{L.}~\bibnamefont{Goldenberg}},\ and\ \bibinfo
  {author} {\bibfnamefont{S.}~\bibnamefont{Wiesner}},\ }%
  \bibfield{journal}{%
  \Doi{10.1103/PhysRevA.54.R1745}{\bibinfo {journal} {Phys. Rev. A}}\ }%
  \textbf{\bibinfo {volume} {54}},\ \bibinfo {pages} {R1745} (\bibinfo {month}
  {Sep}\ \bibinfo {year} {1996}),\
  \url{http://link.aps.org/doi/10.1103/PhysRevA.54.R1745}%
  \bibAnnoteFile{NoStop}{Vaidman1996}%
\bibitem{Erez2004}%
  \BibitemOpen
  \bibfield{author}{%
  \bibinfo {author} {\bibfnamefont{N.}~\bibnamefont{Erez}}, \bibinfo {author}
  {\bibfnamefont{Y.}~\bibnamefont{Aharonov}}, \bibinfo {author}
  {\bibfnamefont{B.}~\bibnamefont{Reznik}},\ and\ \bibinfo {author}
  {\bibfnamefont{L.}~\bibnamefont{Vaidman}},\ }%
  \bibfield{journal}{%
  \Doi{10.1103/PhysRevA.69.062315}{\bibinfo {journal} {Phys. Rev. A}}\ }%
  \textbf{\bibinfo {volume} {69}},\ \bibinfo {pages} {062315} (\bibinfo {month}
  {Jun}\ \bibinfo {year} {2004}),\
  \url{http://link.aps.org/doi/10.1103/PhysRevA.69.062315}%
  \bibAnnoteFile{NoStop}{Erez2004}%
\bibitem{Paz2011}%
  \BibitemOpen
  \bibfield{author}{%
  \bibinfo {author} {\bibfnamefont{G.~A.}\ \bibnamefont{Paz-Silva}}, \bibinfo
  {author} {\bibfnamefont{A.~T.}\ \bibnamefont{Rezakhani}}, \bibinfo {author}
  {\bibfnamefont{J.~M.}\ \bibnamefont{Dominy}},\ and\ \bibinfo {author}
  {\bibfnamefont{D.~A.}\ \bibnamefont{Lidar}},\ }%
  \bibfield{journal}{%
  \Doi{10.1103/PhysRevLett.108.080501}{\bibinfo {journal} {Phys. Rev. Lett.}}\
  }%
  \textbf{\bibinfo {volume} {108}},\ \bibinfo {pages} {080501} (\bibinfo
  {month} {Feb}\ \bibinfo {year} {2012}),\
  \url{http://link.aps.org/doi/10.1103/PhysRevLett.108.080501}%
  \bibAnnoteFile{NoStop}{Paz2011}%
\bibitem{Law2010}%
  \BibitemOpen
  \bibfield{author}{%
  \bibinfo {author} {\bibfnamefont{J.-Q.}\ \bibnamefont{Liao}}\ and\ \bibinfo
  {author} {\bibfnamefont{C.~K.}\ \bibnamefont{Law}},\ }%
  \bibfield{journal}{%
  \Doi{10.1103/PhysRevA.82.053836}{\bibinfo {journal} {Phys. Rev. A}}\ }%
  \textbf{\bibinfo {volume} {82}},\ \bibinfo {pages} {053836} (\bibinfo {year}
  {2010})%
  \bibAnnoteFile{NoStop}{Law2010}%
\bibitem{Gardiner1985}%
  \BibitemOpen
  \bibfield{author}{%
  \bibinfo {author} {\bibfnamefont{C.~W.}\ \bibnamefont{Gardiner}}\ and\
  \bibinfo {author} {\bibfnamefont{M.~J.}\ \bibnamefont{Collett}},\ }%
  \bibfield{journal}{%
  \Doi{10.1103/PhysRevA.31.3761}{\bibinfo {journal} {Phys. Rev. A}}\ }%
  \textbf{\bibinfo {volume} {31}},\ \bibinfo {pages} {3761} (\bibinfo {year}
  {1985})%
  \bibAnnoteFile{NoStop}{Gardiner1985}%
\bibitem{JHS2009}%
  \BibitemOpen
  \bibfield{author}{%
  \bibinfo {author} {\bibfnamefont{J.~H.}\ \bibnamefont{Shapiro}},\ }%
  \bibfield{journal}{%
  \bibinfo {journal} {IEEE. J. Sel. Top. in Quantum Electron.},\ \bibinfo
  {pages} {1547}}%
   (\bibinfo {year} {2009})%
  \bibAnnoteFile{NoStop}{JHS2009}%
\bibitem{Note1}%
  \BibitemOpen
  \bibinfo {note} {Because we consider a one-sided cavity, we should really
  only integrate over $k>0$ in \protect \textrm {Eq.}\protect \tmspace
  +\thinmuskip {.1667em}\protect \textup {\hbox {\mathsurround \z@ \protect
  \normalfont (\ignorespaces \ref {eq:H0ff}\unskip \@@italiccorr )}}. By using
  a truly linear dispersion relation, $\omega _k = ck$, we ensure that the
  unphysical, negative $k$ modes are so far-detuned from the atomic system as
  to be irrelevant.}%
  \bibAnnoteFile{Stop}{Note1}%
\bibitem{Kojima2003}%
  \BibitemOpen
  \bibfield{author}{%
  \bibinfo {author} {\bibfnamefont{K.}~\bibnamefont{Kojima}}, \bibinfo {author}
  {\bibfnamefont{H.~F.}\ \bibnamefont{Hofmann}}, \bibinfo {author}
  {\bibfnamefont{S.}~\bibnamefont{Takeuchi}},\ and\ \bibinfo {author}
  {\bibfnamefont{K.}~\bibnamefont{Sasaki}},\ }%
  \bibfield{journal}{%
  \Doi{10.1103/PhysRevA.68.013803}{\bibinfo {journal} {Phys. Rev. A}}\ }%
  \textbf{\bibinfo {volume} {68}},\ \bibinfo {pages} {013803} (\bibinfo {year}
  {2003})%
  \bibAnnoteFile{NoStop}{Kojima2003}%
\bibitem{Kolner1989}%
  \BibitemOpen
  \bibfield{author}{%
  \bibinfo {author} {\bibfnamefont{B.~H.}\ \bibnamefont{Kolner}}\ and\ \bibinfo
  {author} {\bibfnamefont{M.}~\bibnamefont{Nazarathy}},\ }%
  \bibfield{journal}{%
  \Doi{10.1364/OL.14.000630}{\bibinfo {journal} {Opt. Lett.}}\ }%
  \textbf{\bibinfo {volume} {14}},\ \bibinfo {pages} {630} (\bibinfo {year}
  {1989})%
  \bibAnnoteFile{NoStop}{Kolner1989}%
\bibitem{Kolner1994}%
  \BibitemOpen
  \bibfield{author}{%
  \bibinfo {author} {\bibfnamefont{B.}~\bibnamefont{Kolner}},\ }%
  \bibfield{journal}{%
  \Doi{10.1109/3.301659}{\bibinfo {journal} {IEEE J. of Quantum Electron.}}\ }%
  \textbf{\bibinfo {volume} {30}},\ \bibinfo {pages} {1951 } (\bibinfo {year}
  {1994}),\ ISSN \bibinfo {issn} {0018-9197}%
  \bibAnnoteFile{NoStop}{Kolner1994}%
\bibitem{Kauffman1994}%
  \BibitemOpen
  \bibfield{author}{%
  \bibinfo {author} {\bibfnamefont{M.~T.}\ \bibnamefont{Kauffman}},\ }%
  \emph{\bibinfo {title} {Apllications of Temporal Optical Systems}},\ Ph.D.
  thesis,\ \bibinfo {school} {Stanford University} (\bibinfo {year} {1994})%
  \bibAnnoteFile{NoStop}{Kauffman1994}%
\bibitem{Kuzucu2009}%
  \BibitemOpen
  \bibfield{author}{%
  \bibinfo {author} {\bibfnamefont{O.}~\bibnamefont{Kuzucu}}, \bibinfo {author}
  {\bibfnamefont{Y.}~\bibnamefont{Okawachi}}, \bibinfo {author}
  {\bibfnamefont{R.}~\bibnamefont{Salem}}, \bibinfo {author}
  {\bibfnamefont{M.~A.}\ \bibnamefont{Foster}}, \bibinfo {author}
  {\bibfnamefont{A.~C.}\ \bibnamefont{Turner-Foster}}, \bibinfo {author}
  {\bibfnamefont{M.}~\bibnamefont{Lipson}},\ and\ \bibinfo {author}
  {\bibfnamefont{A.~L.}\ \bibnamefont{Gaeta}},\ }%
  \bibfield{journal}{%
  \Doi{10.1364/OE.17.020605}{\bibinfo {journal} {Opt. Express}}\ }%
  \textbf{\bibinfo {volume} {17}},\ \bibinfo {pages} {20605} (\bibinfo {year}
  {2009})%
  \bibAnnoteFile{NoStop}{Kuzucu2009}%
\bibitem{NC}%
  \BibitemOpen
  \bibfield{author}{%
  \bibinfo {author} {\bibfnamefont{M.~A.}\ \bibnamefont{Nielsen}}\ and\
  \bibinfo {author} {\bibfnamefont{I.~L.}\ \bibnamefont{Chuang}},\ }%
  \emph{\bibinfo {title} {Quantum Computation and Quantum Information}}\
  (\bibinfo {publisher} {Cambridge University Press},\ \bibinfo {year} {2002})%
  \bibAnnoteFile{NoStop}{NC}%
\bibitem{Shen2005}%
  \BibitemOpen
  \bibfield{author}{%
  \bibinfo {author} {\bibfnamefont{J.-T.}\ \bibnamefont{Shen}}\ and\ \bibinfo
  {author} {\bibfnamefont{S.}~\bibnamefont{Fan}},\ }%
  \bibfield{journal}{%
  \Doi{10.1103/PhysRevLett.95.213001}{\bibinfo {journal} {Phys. Rev. Lett.}}\
  }%
  \textbf{\bibinfo {volume} {95}},\ \bibinfo {pages} {213001} (\bibinfo {year}
  {2005})%
  \bibAnnoteFile{NoStop}{Shen2005}%
\bibitem{Nakamura2010}%
  \BibitemOpen
  \bibfield{author}{%
  \bibinfo {author} {\bibfnamefont{O.}~\bibnamefont{Astafiev}}, \bibinfo
  {author} {\bibfnamefont{A.~M.}\ \bibnamefont{Zagoskin}}, \bibinfo {author}
  {\bibfnamefont{A.~A.}\ \bibnamefont{Abdumalikov}}, \bibinfo {author}
  {\bibfnamefont{Y.~A.}\ \bibnamefont{Pashkin}}, \bibinfo {author}
  {\bibfnamefont{T.}~\bibnamefont{Yamamoto}}, \bibinfo {author}
  {\bibfnamefont{K.}~\bibnamefont{Inomata}}, \bibinfo {author}
  {\bibfnamefont{Y.}~\bibnamefont{Nakamura}},\ and\ \bibinfo {author}
  {\bibfnamefont{J.~S.}\ \bibnamefont{Tsai}},\ }%
  \bibfield{journal}{%
  \Doi{10.1126/science.1181918}{\bibinfo {journal} {Science}}\ }%
  \textbf{\bibinfo {volume} {327}},\ \bibinfo {pages} {840} (\bibinfo {year}
  {2010})%
  \bibAnnoteFile{NoStop}{Nakamura2010}%
\end{thebibliography}%
\end{document}